\documentclass[preprint,copyright,creativecommons]{eptcs}
\providecommand{\event}{PLACES 2025} 

\usepackage[utf8]{inputenc}
\usepackage[T1]{fontenc}
\usepackage[dvipsnames]{xcolor}
\usepackage{amsmath,amssymb,stmaryrd,amsthm}
\usepackage[final]{listings}
\usepackage{mathpartir}
\usepackage{xspace}
\usepackage{tikz}
\usepackage{wrapfig}
\usepackage{hyperref}
\usepackage[noabbrev]{cleveref}
\usepackage[inline]{enumitem}
\usepackage{appendix} 
\usepackage{pifont}%
\usepackage{adjustbox}
\usepackage{array}
\usepackage{lipsum}
\usepackage{multicol}

\usepackage{iftex}

\ifpdf
  \usepackage{underscore}         
  \usepackage[T1]{fontenc}        
\else
  \usepackage{breakurl}           
\fi

\title{Local Type Inference for Context-Free Session Types}
\author{
Bernardo Almeida \qquad\qquad Andreia Mordido \qquad\qquad Vasco T. Vasconcelos
\institute{LASIGE, Faculty of Sciences, University of Lisbon, Portugal}
\email{\{bpdalmeida,afmordido,vmvasconcelos\}@ciencias.ulisboa.pt}
}

\newcommand{\titlerunning}{Local type inference}
\newcommand{\authorrunning}{B. Almeida, A. Mordido \& V.T. Vasconcelos}

\hypersetup{
  bookmarksnumbered,
  pdftitle    = {\titlerunning},
  pdfauthor   = {\authorrunning},
  pdfsubject  = {EPTCS},               
}

\newcommand{\freest}{\textsc{FreeST}\xspace}

\newcommand{\kindc}[1]{\color{ForestGreen}{#1}} 
\newcommand{\typec}[1]{\begingroup\color{RoyalBlue}{#1}\endgroup}
\newcommand{\termc}[1]{\color{RedOrange}{#1}}
\newcommand{\blk}[1]{\color{black}{#1}}

\newcommand{\kw}[1]{\mathsf{#1}}
\newcommand{\syntax}[1]{\mathtt{#1}}

\newcommand{\grmeq}{\; ::= \;\;}
\newcommand{\grmor}{\;\mid\;}

\newcommand{\multstyle}[1]{\kindc{\textbf{\oldstylenums{#1}}}}

\newcommand\symK{\kindc{\kappa}}

\newcommand\lin{\multstyle{1}}
\newcommand\un{\multstyle{*}}

\newcommand{\kind}{\symK}

\newcommand\polarity{\typec{\sharp}}
\newcommand\polOut{\typec{!}}
\newcommand\polIn{\typec{?}}
\newcommand\view{\typec{\star}}
\newcommand\viewOut{\typec{\oplus}}
\newcommand\viewIn{\typec{\&}}

\newcommand\choice[1][\cdot]{\typec{\{{#1}\}}}

\newcommand{\subs}[3]{{\typec{#3}}[{\typec{#1}}/{\typec{#2}}]}

\newcommand\ifthenelse[3]{\text{if}\;\, #1 \;\,\text{then}\;\, #2 \;\,\text{else}\;\, #3}
\newcommand{\isClosures}{\syntax{isAbs}}
\newcommand{\isClosure}[3]{\ifthenelse{\isClosures \;#1}{#2}{#3}}
\newcommand{\muredex}[1][\TRec\TVar\kind\type]{\mu\text{-redex}(#1)}
\newcommand{\muredexPrime}[1][\TRec\TVar\kind\type]{\mu\text{-redex'}(#1)}

\newcommand\type[1][\symT]{\typec{#1}}
\newcommand\typeI[2][\symT]{\typec{#1_{\!\!{#2}}}}

\newcommand\TInt{\typec{\kw{Int}}}
\newcommand\TBool{\typec{\kw{Bool}}}

\newcommand\TSkip{\typec{\kw{Skip}}}
\newcommand\TWait{\typec{\kw{Wait}}}
\newcommand\TClose{\typec{\kw{Close}}}
\newcommand\TEnd[1][\polarity]{\typec{\kw{End}_{#1}}}
\newcommand{\TMsg}[2][\polarity]{{#1}\,\typec{#2}}

\newcommand{\TChoice}[4][\view]{{#1}\,\typec{\choice[#2\colon#3_{#2}]_{#2\in#4}}}

\newcommand{\TSemi}[2]{\typec{{#1};{#2}}}
\newcommand{\TFun}[3][\mult]{\typec{{#2}\rightarrow{#3}}}

\newcommand{\TAll}[3]{\typec{\forall{\typec{#1}}.\typec{#3}}}
\newcommand{\TRec}[3]{\typec{\mu{\typec{#1}}.{\typec{#3}}}}
\newcommand{\TVar}[1][\alpha]{\typec{#1}}
\newcommand{\IVar}[1][X]{\typec{#1}}

\newcommand{\expr}[1][\symE]{\termc{#1}}

\newcommand\EVar[1][x]{\expr[#1]}

\newcommand\EAbs[4][\mult]{\expr[\lambda{#2}\colon{\typec{#3}}.{\expr[#4]}]}
\newcommand\ETAbs[3]{\expr[\Lambda{\typec{#1}}.{\expr[#3]}]}

\newcommand\EApp[2]{\expr[#1\;#2]}
\newcommand\EArgs[1][\pi]{\expr[#1]}
\newcommand\EAppHead[1][h]{\expr[#1]}

\newcommand{\infrule}[3]{\inferrule* [lab=#1]{#2}{#3}}
\newcommand{\axiom}[2]{\infrule {#1}{}{#2}}

\newcommand{\declrel}[2]{\emph{#1}\hfill\fbox{{#2}}}

\newcommand{\judgementrel}[3]{{#1} \; {#2} \; {#3}}

\newcommand{\judgementrelctx}[4]{{#1} \vdash \judgementrel{#2}{#3}{#4}}

\newcommand{\Empty}{\varnothing}

\newcommand\fresh{\,\operatorname{fresh}}
\newcommand\TypeEquiv{\simeq}

\newcommand\undefined[1]{{#1}\text{ undefined}}

\newcommand{\synthetise}{\Rightarrow}

\newcommand{\isExprAlg}[5][\Delta]{\judgementrelctx{#2}{#3}{\synthetise}{{#4}\mid{#5}}}
\newcommand{\isExprAlgAgainst}[5][\Delta]{\judgementrelctx{#2}{{#3}\colon{#4}}{\synthetise}{#5}}

\newcommand{\unify}[4][\Xi]{\judgementrelctx{#1}{#2}{\dot{=}}{#3} \leadsto {#4}}

\newcommand\baseInstOut[5][\Gamma]{{#1}\vdash_\textsc{inst}{#2\,;\,#3} \rightsquigarrow \judgementrel{#4}{;}{#5} }
\newcommand\baseInstIn[6][\Gamma]{{#1}\vdash_\textsc{I}{#2\,;\,#3} \rightsquigarrow \judgementrel{#4\,;\,#5}{;}{#6} }
\newcommand\quicklook[4][\Gamma]{{#1} \vdash_\textsc{QL} {#2} \colon \judgementrel{#3}{\rightsquigarrow}{#4} }

\newcommand{\reduce}[2]{\judgementrel{#1}{\rightarrow}{#2}}
\newcommand\fiv[1]{\text{fiv}(#1)}

\newcommand{\rulename}[2]{\textsc{\small{#1}-{#2}}\xspace}

\newcommand{\rulenameReduction}[1]{\rulename{R}{#1}}
\newcommand\ruleRSkip{\rulenameReduction{Skip}}
\newcommand\ruleRAssoc{\rulenameReduction{Assoc}}
\newcommand\ruleRSemi{\rulenameReduction{Semi}}
\newcommand\ruleRDistrib{\rulenameReduction{Distrib}}
\newcommand\ruleRRec{\rulenameReduction{Rec}}

\newcommand{\rulenameInst}[1]{\rulename{I}{#1}}
\newcommand\ruleIOut{\rulenameInst{Outer}}
\newcommand\ruleIRes{\rulenameInst{Result}}
\newcommand\ruleIAll{\rulenameInst{AllExp}}
\newcommand\ruleITArg{\rulenameInst{AllType}}
\newcommand\ruleIArg{\rulenameInst{Arg}}
\newcommand\ruleIVar{\rulenameInst{Var}}

\newcommand{\rulenameQL}[1]{\rulename{QL}{#1}}
\newcommand\ruleQLApp{\rulenameQL{App}}
\newcommand\ruleQLOther{\rulenameQL{Other}}

\newcommand{\rulenameM}[1]{\rulename{M}{#1}}
\newcommand\ruleMfiv{\rulenameM{fiv}}
\newcommand\ruleMRedex{\rulenameM{redex}}
\newcommand\ruleMReduceL{\rulenameM{Reduce-L}}
\newcommand\ruleMReduceR{\rulenameM{Reduce-R}}
\newcommand\ruleMVarL{\rulenameM{IVar-L}}
\newcommand\ruleMVarR{\rulenameM{IVar-R}}
\newcommand\ruleMSkip{\rulenameM{Skip}}
\newcommand\ruleMEnd{\rulenameM{End}}

\newcommand\ruleMVar{\rulenameM{Var}}
\newcommand\ruleMSemi{\rulenameM{Semi}}
\newcommand\ruleMChoice{\rulenameM{Choice}}
\newcommand\ruleMMsg{\rulenameM{Msg}}
\newcommand\ruleMArrow{\rulenameM{Arrow}}

\newcommand\ruleMAll{\rulenameM{All}}
\newcommand\ruleMSemiL{\rulenameM{Semi-L}}
\newcommand\ruleMSemiR{\rulenameM{Semi-R}}

\newcommand{\rulenameCA}[1]{\rulename{$\Downarrow$}{#1}}
\newcommand\ruleCApp{\rulenameCA{App}}

\newcommand{\rulenameS}[1]{\rulename{$\Uparrow$}{#1}}
\newcommand\ruleSApp{\rulenameS{App}}

\definecolor{amber}{rgb}{1.0, 0.75, 0.0}
\definecolor{greenTick}{rgb}{0.0, 0.5, 0.0}

\newcolumntype{R}[2]{%
    >{\adjustbox{angle=#1,lap=\width-(#2)}\bgroup}%
    l%
    <{\egroup}%
}

\newcommand{\etal}{et al.\xspace} 

\newcommand\Small{\small}

\definecolor{darkviolet}{rgb}{0.5,0,0.4}
\definecolor{darkgreen}{rgb}{0,0.4,0.2}
\definecolor{darkblue}{rgb}{0.1,0.1,0.9}
\definecolor{darkgrey}{rgb}{0.5,0.5,0.5}
\definecolor{lightblue}{rgb}{0.4,0.4,1}

\lstdefinestyle{eclipse}{
  breaklines=true,
  basicstyle=\sffamily\Small,
  emphstyle=\color{red}\bfseries,
  keywordstyle=\color{darkviolet}\bfseries,
  commentstyle=\color{darkgreen},
  stringstyle=\color{darkblue},
  numberstyle=\color{darkgrey},
  emphstyle=\color{red},
  showstringspaces=false,
}

\begin{document}
\maketitle

\begin{abstract}
  We address the problem of local type inference for a language based on System
  F with context-free session types. We present an algorithm that leverages the
  bidirectional type checking approach to propagate type information, enabling
  first class polymorphism while addressing the intricacies brought about by the
  sequential composition operator and type equivalence. The algorithm improves
  the language's usability by eliminating the need for type annotations at type
  application sites.
\end{abstract}

\section{Introduction}



Type inference is a fundamental aspect of programming language design, allowing
type information to be automatically determined. This mechanism simplifies code
development, while enhancing code readability. By inferring types, compilers
gather enough type information to statically verify programs, thereby preventing a
range of runtime errors.
However, full type inference in expressive type systems such as System F present
considerable theoretical and practical challenges. The presence of impredicative
polymorphism, where type variables can be instantiated with polymorphic types, 
increases the complexity of type inference, leading to undecidability of
type checking in the general case~\cite{Wells94}. Therefore, type systems based
on System F usually impose syntactic constraints via type annotations.

Research in this area continues to explore strategies for balancing decidability
with the amount of annotations required. While excessive annotations, such as
those required in polymorphic applications, can be cumbersome and contribute to
code complexity, annotations in top-level function definitions and their bound
variables are valuable for documentation and enhance code clarity.
Local type inference was proposed by Pierce and Turner~\cite{PierceT98} aiming
at infering type annotations at application sites. Furthermore, due to its
locality (only at the application level) these mechanisms offer better error
messages to the programmer than those provided by full type inference.

In this work we propose a local type inference algorithm for
\freest~\cite{freest, AlmeidaMV19}, a concurrent programming language based on
System F where processes communicate on heterogeneously typed-channels governed
by context-free session types~\cite{ThiemannV16}. Context-free session types are
able to describe non-regular protocols and include types for describing message
sending (\lstinline|!T|) and receiving (\lstinline|?T|), sequential composition
of two protocols (\lstinline|T;U|) and the neutral element of composition
\lstinline|Skip|. The sequential composition operator is particularly effective
in protocol composition and decomposition. The gain in expressivity comes with a
cost: type equivalence, usually defined by a (non-necessarily finite) type
bisimulation, presents a substantial challenge. Consequently (local) type
inference becomes slightly more complicated than usual since we need to account
for a complex notion of type equivalence.


Consider as an example the following function that takes an integer value and a
channel, sends the successor of the value on the channel, and returns the
continuation of that channel.
\begin{lstlisting}
f : Int -> !Int;Close -> Close
f x c = send (x + 1) c 
\end{lstlisting}
The channel is of type \lstinline|!Int;Close|, meaning that an integer
should be sent and then the channel should be closed.
Function \lstinline|f| consumes the initial part of the channel (the
\lstinline|!Int| part) and returns the channel (now of type \lstinline|Close|).
The primitive \lstinline|send| function has type
\lstinline|foralla . a -> forallb . !a;b -> b|, meaning that, whenever we use
it, we start by instantiating type variable \lstinline|a| with, say, type
\lstinline|T|, then provide a value of that type, then instantiate type variable
\lstinline|b| with, say, type \lstinline|S|, and finally provide a channel of
type \lstinline|!T;S|. In return, we expect the continuation of the channel (at
this stage of type \lstinline|S|). In practice, the programmer must provide
explicit annotations to guide the type checker. In \freest, type instantiations
are denoted by \lstinline|@|. In order to properly type check, our function
\lstinline|f| should thus be written as follows:
\begin{lstlisting}
f : Int -> !Int;Close -> Close
f x c = send @Int (x + 1) @Close c 
\end{lstlisting}

To be faithful to the type specified for \lstinline|f|, the programmer
instantiates type variable \lstinline|a| with type \lstinline|Int| (through
annotation \lstinline|@Int|) and \lstinline|b| with type \lstinline|Close|
(through annotation \lstinline|@Close|). After all, the programmer has to figure
out not only \emph{which annotations} are required but also \emph{where to place
  them}. In such a small example the task may seem easy, but with more complex
protocols things escalate quickly. One example is when we require polymorphic
instantiation, that is, the annotation is itself a polymorphic type.

Aiming to get rid of this burden of annotations in type 
applications, we propose a new local type inference for \freest. 
Our proposal builds primarily on Quick Look~\cite{SerranoHJV20}, which enables
the inference of type annotations in polymorphic applications. However, the
presence of context-free session types introduces additional subtleties. In
particular, we must account for explicit recursive types and the monoidal laws
introduced by the sequential composition operator (with type \lstinline|Skip| as
the identity element), and treats \lstinline|Close| and \lstinline|Wait| as left
absorbing types.

The main contributions of our work are:
\begin{itemize}
    \item A local type inference algorithm for \freest,
    \item A novel type matching algorithm that deals with context-free session types,
    \item A prototype implementation, integrated in the \freest compiler.  
\end{itemize}

The rest of the paper is organised as follows. \Cref{sec:syntax} introduces a
rather stripped down version of \freest, yet rich enough to explain the main
issues in local type inference for context-free session types. Then,
\cref{sec:local-typer-inference} introduces the inference algorithm.
\Cref{sec:results} evaluates the algorithm, \cref{sec:related-work} discusses
related work, and \cref{sec:future} concludes the paper.





\section{Syntax} 
\label{sec:syntax}

This section briefly introduces the language we use to illustrate local type
inference.
We work with a rather stripped down version of the \freest programming
language~\cite{AlmeidaMTV22}, keeping only the relevant constructors. 
The language relies on a few base sets: \emph{type variables}, denoted by
$\TVar, \TVar[\beta], \TVar[\gamma]$, \emph{term variables} denoted by
$\EVar, \EVar[y], \EVar[z]$ and, anticipating type inference,\emph{
  instantiation variables} denoted by $\IVar, \IVar[Y], \IVar[Z]$. The syntax is
in~\cref{fig:syntax}.

\begin{figure}[t!]
  \begin{align*}
    \polarity \grmeq & \polOut \grmor \polIn                      &\text{Polarities}\\
    \view \grmeq & \viewIn \grmor \viewOut                        &\text{Views}\\
    \type \grmeq & \TEnd
                   \grmor \TMsg\type
                   \grmor \TChoice \ell TL 
                   \grmor \TSemi\type\type
                   \grmor \TSkip 
                   \grmor \TRec\TVar\kind\type &\text{Types}\\
                   &\grmor \TVar
                    \grmor \TFun\type\type
                     \grmor \TAll\TVar\kind\type
                   \grmor \IVar                                         
    \\
%
    \expr \grmeq & \EAbs x\type\expr
          \grmor \ETAbs\TVar\kind\expr
          \grmor \EApp\EAppHead{\EArgs_1\blk,\dots,\EArgs_n}
    & \text{Expressions}
    \\
    \EAppHead \grmeq & \EVar \grmor \EAbs x\type\expr
          \grmor \ETAbs\TVar\kind\expr & \text{Application heads}
    \\
    \EArgs \grmeq & \expr \grmor \type & \text{Function arguments}
  \end{align*}
  \caption{Syntax of types and expressions}
  \label{fig:syntax}
\end{figure}


%


\emph{Session types} include channel closing ($\TClose$ and $\TWait$,
collectively denoted by $\TEnd$), message sending ($\TMsg[\polOut]\type$) and
receiving ($\TMsg[\polIn]\type$), internal ($\TChoice[\viewOut] \ell T L$) and
 external ($\TChoice[\viewIn] \ell T L$) choices,
the sequential composition of types ($\TSemi{\typeI 1}{\typeI 2}$) and $\TSkip$,
denoting the absence of communication. 
%
\emph{Functional types} include functions ($\TFun[\lin]{\typeI 1}{\typeI
  2}$) 
and universal types ($\TAll\TVar\kind\type$). Pairs and sums, records and
variants, the unit type and other standard functional types can be easily
incorporated.
Recursive types ($\TRec\TVar\kind\type$) are sessions only, for simplicity.
%

Types are filtered by a collection of type formation rules that essentially
guarantee contractiveness, that is, that guarantee that continuous unfolding of
recursive types eventually yields a proper (non-$\typec\mu$) type constructor.
The situation is slightly complicated by the introduction of context-free
session types: $\TSkip$ is not considered a proper type constructor for this
purpose, so that $\TRec\alpha{\kind}{\TSemi\TSkip\alpha}$ is not a well formed
type.
The details are in Almeida \etal~\cite{AlmeidaMTV22}.

\emph{Expressions} include term abstraction $\EAbs\EVar\type\expr$ and type
abstraction $\ETAbs\TVar\kind\expr$. In this work applications are represented
as \textit{n-ary} constructs, denoted by $\EApp\EAppHead{\overline\EArgs}$,
where application heads $\EAppHead$ is either term variables, term abstractions,
or type abstractions. The arguments $\expr[\overline\EArgs]$ may consist of
either expressions or types. Term variables are simply applications with an
empty list of arguments. This generalisation provides for a unified treatment
of the traditional term and type application, and is widely adopted in type
inference algorithms to ensure that each application node carries maximal
information~\cite{PierceT98,SerranoHJV20}.
%


\section{Local type inference}
\label{sec:local-typer-inference}

Local type inference for session types requires a few novel notions that we now
introduce.

\emph{Type reduction}, defined in the top of~\cref{fig:type-reduction}, is a partial
function defined on session types. It performs one-step reduction, while
exploring the monoidal semantics of sequential composition. Rule \ruleRSkip
eliminates $\TSkip$, the neutral element of sequential composition. Rule
\ruleRAssoc enforces the monoidal associativity law, reducing
$\TSemi{(\TSemi{\typeI 1}{\typeI 2})}{\typeI 3}$ to
$\TSemi{\typeI 1}{(\TSemi{\typeI 2}{\typeI 3})}$. Rule \ruleRSemi reduces the
first element of a sequential composition and combines the result with the
second element. Rule \ruleRDistrib distributes a sequential composition over a
choice. Rule \ruleRRec unfolds a recursive type.

Let $\type$ be the type
$\TRec\TVar\kind{\TSemi{(\TSemi{\TMsg[\polOut]\TInt}{\TMsg[\polIn]\TBool})}{\TVar}}$.
We can observe that after two reduction steps we expose the first proper
constructor, namely $\TMsg[\polOut]\TInt$. Indeed, through rules
\ruleRRec and \ruleRAssoc we get:
$\type \rightarrow
\TSemi{(\TSemi{\TMsg[\polOut]\TInt}{\TMsg[\polIn]\TBool})}\type \rightarrow
\TSemi{\TMsg[\polOut]\TInt}{(\TSemi{\TMsg[\polIn]\TBool}\type)}$. This process
will become important when comparing two different types for matching: 
notice that based on their \emph{behaviour}, both
$\type$ and $\TSemi{\TMsg[\polOut]\TInt}{(\TSemi{\TMsg[\polIn]\TBool}\type)}$
express the same communication and thus must match against each other.

Another important (and novel) concept is that of \emph{$\mu$-redex}, 
defined in the bottom of~\cref{fig:type-reduction}.
Conceptually, the $\mu$-redex of a recursive type is the recursive type itself.
Given the semantics of sequential composition, the $\mu$-redex of a recursive
type $\type$ composed with something else is also $\type$. Therefore, the types
$\TSemi{(\TRec\TVar\kind
  {\TSemi{\TMsg[\polOut]\TInt}{\TVar}})}{\TMsg[\polIn]\TBool}$ and
$\TRec\TVar\kind {\TSemi{\TMsg[\polOut]\TInt}{\TVar}}$ have a common
$\mu$-redex, namely $\TRec\TVar\kind {\TSemi{\TMsg[\polOut]\TInt}{\TVar}}$.
The function $\mu$-redex is total: 
it either returns a singleton set containing the redex of the given
type $\type$ or the empty set $\Empty$ if the auxiliary function $\mu$-redex' is
undefined on $\type$. Keeping track of the $\mu$-redexes 
is
crucial to ensure termination of type matching, as we clarify below.

\begin{figure}[t!]
  \declrel{Type reduction}{$\reduce{\type_{\text{in}}}{\type_{\text{out}}}$}
  \begin{gather*}
    \axiom{\ruleRSkip}{\reduce{\TSemi\TSkip\type}\type}
    \quad
    \axiom{\ruleRAssoc}{\reduce {\TSemi{(\TSemi{\typeI 1}{\typeI 2})}{\typeI 3}}
      {\TSemi{\typeI 1}{(\TSemi{\typeI 2}{\typeI 3})}}
    }
    \quad
    \infrule{\ruleRSemi}
    {\reduce{\typeI 1}{\typeI 3}}
    {\reduce{\TSemi{\typeI 1}{\typeI 2}}{\TSemi{\typeI 3}{\typeI 2}}}
    \\
    \axiom{\ruleRDistrib}{\reduce{\TSemi{\TChoice \ell TL }{\typeI 1}}
      {\typec{\view\choice[\ell\colon{\TSemi{T_{\ell}}{\typeI 1}}]_{\ell\in L}} }
    }
    \quad
    \axiom{\ruleRRec} {\reduce{\TRec\TVar\kind\type}
      {\subs{\TRec\TVar\kind\type}{\TVar}{\type}}}
  \end{gather*}
  \declrel{$\mu$-redex}{$\muredex[\type] = \{\type\}$}
  \vspace{-4ex}
  \begin{multicols}{2}
    \begin{equation*}
    \muredex[\typeI 1] =
    \begin{cases}
      \{\typeI 2\} & \text{if }\muredexPrime[\typeI 1] = {\typeI 2} \\
      \Empty & \text{if } \undefined{\muredexPrime[{\typeI 1}]}\\
    \end{cases}
  \end{equation*}
  
  \begin{align*}
    \muredexPrime &= \TRec\TVar\kind\type
    \\
    \muredexPrime [\TSemi{(\TRec\TVar\kind{\typeI 1})}{\typeI 2}]
    &= \TRec\TVar\kind{\typeI 1}
  \end{align*}     
  \end{multicols}
  \caption{Type reduction and $\mu$-redexes}
  \label{fig:type-reduction}
  \label{fig:muredex}
\end{figure}


Equipped with type reduction and $\mu$-redexes, we can introduce \emph{type
  matching}, which plays a central role in the inference process. This is the
major deviation from the original work by Serrano \etal \cite{SerranoHJV20}
given that we need to deal with both the monoidal laws and the left absorbing
elements.
%
In general, to ensure that types $\TFun[\un]\TInt\IVar$ and
$\TFun[\un]\TInt\TBool$ match, it must the case that the instantiation variable
$\IVar$ must be equal to $\TBool$. This simple example poses no challenge, but
the same does not happen with the recursive types and sequential composition.
Consider the types
$\typec{S_1} = \TRec\TVar\kind{(\TSemi{(\TSemi {\TMsg[\polOut]\TInt}\TVar)}
  \IVar)}$ and
$\typec{S_2} = \TRec{\TVar[\beta]}\kind{(\TSemi
  {\TMsg[\polOut]\TInt}{\TVar[\beta]})}$. It is not evident what is the value of
$\IVar$ or even if the two types match. We introduce a matching algorithm
capable of dealing with such types.

\begin{figure}[t!]
  \declrel{Type matching}{$\unify[\Xi_{\text{in}}]{\type_{\text{in}}}{\type_{\text{in}}}{\Theta_{\text{out}}}$}
  \begin{gather*}
    \infrule{\ruleMfiv}
      {\fiv{\typeI 1, \typeI 2} = \Empty}
      {\unify{\typeI 1}{\typeI 2}{\Empty}}
    \quad
    \infrule{\ruleMRedex}
      {\muredex[\typeI 1, \typeI 2] \subseteq \Xi}
      {\unify{\typeI 1}{\typeI 2}{\{\fiv{\typeI 1, \typeI 2} \mapsto \TSkip \}}}
    \quad
    \infrule{\ruleMReduceL}
    {\muredex[\typeI 1] \cap \Xi = \Empty
      \quad \reduce{\typeI 1}{\typeI 3}\\\\
      \unify[\Xi,{\muredex[\typeI 1]}]{\typeI 3}{\typeI 2}{\Theta}
    }
    {\unify{\typeI 1}{\typeI 2}{\Theta}}
    \\
    \infrule{\ruleMReduceR}
    {\muredex[\typeI 2] \cap \Xi = \Empty
      \quad \reduce{\typeI 2}{\typeI 3}
      \\\\ \unify[\Xi,{\muredex[\typeI 2]}]{\typeI 1}{\typeI 3}{\Theta}
    }
      {\unify{\typeI 1}{\typeI 2}{\Theta}}
    \quad
    \axiom{\ruleMVarL}{\unify\IVar\type\{\IVar \mapsto \type\}}
    \quad
    \axiom{\ruleMVarR}{\unify\type\IVar\{\IVar \mapsto \type\}}
    \\
    \axiom{\ruleMSkip}{\unify\TSkip\TSkip\Empty}
    \quad
    \axiom{\ruleMEnd}{\unify\TEnd\TEnd\Empty}
    %
    %
    \quad
    \axiom{\ruleMVar}{\unify\TVar\TVar\Empty}
    \quad
    \infrule{\ruleMMsg}
      {\unify{\typeI 1}{\typeI 2}\Theta}
      {\unify{\TMsg{\typeI 1}}{\TMsg{\typeI 2}}\Theta}
    \\
    \infrule{\ruleMSemiL}
            {\unify{\typeI 1}{\typeI 3}{\Theta_1} \quad
             \unify{\typeI 2}{\TSkip}{\Theta_2}  }
            {\unify{\TSemi{\TMsg{\typeI 1}}{\typeI 2}}{\TMsg{\typeI 3}}{\Theta_1\circ\Theta_2}}
    \quad    
    \infrule{\ruleMSemiR}
            {\unify{\typeI 1}{\typeI 2}{\Theta_1} \quad
             \unify{\TSkip}{\typeI 3}{\Theta_2}  }
            {\unify{\TMsg{\typeI 1}}{\TSemi{\TMsg{\typeI 2}}{\typeI 3}}{\Theta_1\circ\Theta_2}}
    \\ 
    \infrule{\ruleMSemi}
      {\unify{\typeI 1}{\typeI 3}{\Theta_1} \quad \unify{\Theta_1\typeI 2}{\Theta_1\typeI 4}{\Theta_2}}
      {\unify{\TSemi{\typeI 1}{\typeI 2} }{\TSemi{\typeI 3}{\typeI 4}}{\Theta_1\circ\Theta_2}}
    \quad
    \infrule{\ruleMChoice}
      {\unify{\typeI \ell}{\typeI[T'] \ell}{\Theta_\ell} \quad(\forall \ell\in L)}
      {\unify{\TChoice \ell TL}{\TChoice \ell {T'}L}{\circ\Theta_\ell}} 
    \\
    \infrule{\ruleMArrow}
      {\unify{\typeI 1}{\typeI 3}{\Theta_1}\quad \unify{\Theta_1\typeI 2}{\Theta_1\typeI 4}{\Theta_2}}
      {\unify{\TFun{\typeI 1}{\typeI 2}}{\TFun{\typeI 3}{\typeI 4}}{\Theta_1\circ\Theta_2}}
    \quad
    %
    %
    \infrule{\ruleMAll}
            {\typec\gamma\fresh\quad
              \unify{\subs\gamma\alpha{\typeI 1}}{\subs\gamma\beta{\typeI 2}}\Theta}
            {\unify{\TAll\TVar\kind{\typeI 1}}{\TAll{\TVar[\beta]}\kind{\typeI 2}}\Theta}
  \end{gather*}
  \caption{Type matching}
  \label{fig:type-matching}
\end{figure}


The judgment for type matching is of the form
$\unify{\typeI 1}{\typeI 2}{\Theta}$ and reads as ``match types ${\typeI 1}$ and
${\typeI 2}$ under the set $\Xi$ containting the $\mu$-redexes of the visited
types and produce a substitution $\Theta$.'' The rules, in
\cref{fig:type-matching}, have an algorithmic reading when tried in order of
presentation.
Apart from function $\mu$-redex, type matching uses function $\fiv\type$
that yields the free instantiation variables of $\type$.
If none of the types under consideration contain instantiation variables, then
the result is the empty set (rule \ruleMfiv).

Rule \ruleMRedex returns a mapping between the remaining instantiation variables
and $\TSkip$ if the $\mu$-redexes of $\typeI 1$ and $\typeI 2$ were visited
(that is, if they appear in $\Xi$). Recall types $\typec{S_1} = \TRec\TVar\kind{(\TSemi{(\TSemi {\TMsg[\polOut]\TInt}\TVar)}
  \IVar)}$ and $\typec{S_2} = \TRec{\TVar[\beta]}\kind{(\TSemi
  {\TMsg[\polOut]\TInt}{\TVar[\beta]})}$. 
If $\typec{S_1}$'s redex is in $\Xi$, we argue that, semantically, it makes sense to substitute
$\IVar$ with $\TSkip$ since it will always be unreachable.

Rules \ruleMReduceL and \ruleMReduceR are similar (one for each side of the
equation). We assume that both $\typeI 1$ and $\typeI 2$ are well formed. In
particular, we assume recursive types to be contractive, as explained in~\cref{sec:syntax}. 
With that in mind we can assume that after a
finite number of reduction steps (\cref{fig:type-reduction}), some constructor
will be exposed.
If the $\mu$-redex is not in $\Xi$, then we perform a one-step reduction and
continue recursively with the contractum, while adding the redex to $\Xi$.

Axioms \ruleMVarL and \ruleMVarR match a type against an instantiation variable
and return that matching in the form of a substitution. The axioms \ruleMSkip,
\ruleMEnd and \ruleMVar return the empty substitution. Rule \ruleMMsg matches
the message type $\typeI 1$ against the message type $\typeI 2$.
Rules \ruleMSemiL and \ruleMSemiR account for session continuations on just one
side of the equation and therefore should match with $\TSkip$. Rules \ruleMSemi,
\ruleMChoice and \ruleMArrow are similar to rule \ruleMMsg. Rule \ruleMAll
generates a fresh type variable $\IVar[\gamma]$ and substitutes the bound variables in each
side of the equation, continuing with the bodies of the types.


Now let us consider $\typec{S_1}$ and $\typec{S_2}$ introduced above. Do they match?
Which value should $\IVar$ be assigned to?
Let us try to apply the rules from \cref{fig:type-matching} in order starting with
$\Xi = \Empty$. The result of applying the rules (shown in the first column) is
in \cref{fig:match-example}.

\begin{figure}[t!]
  \centering
  \begin{equation*}
    \begin{array}{c c c c c}
      &\typec{S_1} = \TRec\TVar\kind{(\TSemi{(\TSemi {\TMsg[\polOut]\TInt}\TVar)} \IVar)} & \dot{=} & \typec{S_2} = \TRec{\TVar[\beta]}\kind{(\TSemi {\TMsg[\polOut]\TInt}{\TVar[\beta]})} & \Xi = \Empty \\
      &\downarrow && \downarrow&\\
      \ruleMReduceL & \TSemi {(\TSemi {\TMsg[\polOut]\TInt}{\typec{S_1}})} \IVar & \dot{=} & \typec{S_2} & \Xi = \typec{S_1} \\
      &\downarrow && \downarrow&\\
      \ruleMReduceL & \TSemi {\TMsg[\polOut]\TInt} {(\TSemi {\typec{S_1}}{\IVar})} & \dot{=} & \typec{S_2} & \Xi = \typec{S_1} \\
      &\downarrow && \downarrow&\\
      \ruleMReduceR & \TSemi {\TMsg[\polOut]\TInt} {(\TSemi {\typec{S_1}}{\IVar})} & \dot{=} & \TSemi {\TMsg[\polOut]\TInt}{\typec{S_2}} & \Xi = \typec{S_1}, \typec{S_2} \\
      &\downarrow && \downarrow&\\
      \ruleMSemi & \TMsg[\polOut]\TInt & \dot{=} & \TMsg[\polOut]\TInt & \checkmark \\
      \ruleMSemi & \TSemi{\typec{S_1}}\IVar & \dot{=} & \typec{S_2} & \Xi = \typec{S_1}, \typec{S_2} \\
      &\downarrow && \downarrow&\\
      \ruleMRedex & \muredex[\TSemi{\typec{S_1}} \IVar] = \{\typec{S_1}\} \subseteq \Xi && \muredex[\typec{S_2}] = \{\typec{S_2}\} \subseteq \Xi&\checkmark
    \end{array}
  \end{equation*}
  \caption{Running the match algorithm on types $\TRec\TVar\kind{(\TSemi{(\TSemi {\TMsg[\polOut]\TInt}\TVar)} \IVar)}$ and $\TRec{\TVar[\beta]}\kind{(\TSemi {\TMsg[\polOut]\TInt}{\TVar[\beta]})}$}
  \label{fig:match-example}
\end{figure}

The resulting substitution is $\Theta = \{\IVar\mapsto\TSkip\}$, because the
variable will never be reached. With this example we highlighted the need for
type reduction (rules \ruleMReduceL and \ruleMReduceR) and the need to record
the $\mu$-redexes. Without keeping the visited $\mu$-redices, types would be
reducing eternally without being able to be matched.

Basic instantiation, in \cref{fig:base-inst}, adapted from Serrano
\etal~\cite{SerranoHJV20}, transforms a polymorphic type into a monomorphic type
by replacing each polymorphic variable with an instantiation variable---denoted
by $\IVar, \IVar[Y], \IVar[Z]$---for each $\forall$-quantifier.
Judgment $\baseInstOut {\typeI 1}{\overline\EArgs}{\typec{\overline U}}{\typeI 2}$
reads ``instantiate type $\typeI 1$ guided by function arguments
$\overline\EArgs$ and return the types of each term argument $\typec{\overline U}$
and the function's return type $\typeI 2$''. Internally, it uses the judgment
$\baseInstIn {\typeI 1}{\overline\EArgs}\Theta{\typec{\overline U}}{\typeI 2}$ which
also produces a substitution $\Theta$ that keeps the result of unifying each
argument in $\typec{\overline U}$.

Rule \ruleIRes applies when the list of arguments
$\overline\EArgs$ is empty, returning $\type$ as the result. Rules \ruleIAll and
\ruleITArg handle polymorphic types: \ruleIAll is applied with value arguments,
replacing the bound variable with a fresh instantiation variable, while
\ruleITArg replaces it with the type argument $\typeI 2$.
Rule \ruleIArg applies when the type is a function and the argument is an
expression $\expr$. It analyses the argument by either traversing a nested
application or matching an instantiation variable with a type. The resulting
substitution is applied to both $\typeI 1$ (added to $\typec{\overline U}$) and
$\typeI 2$ (instantiated recursively).
Rule \ruleIVar applies when there are remaining arguments and the type is an
instantiation variable. In this case, the type must represent a function,
captured by the substitution $\Theta_1 = [\IVar :=
\TFun[\un]{\IVar[Y]}{\IVar[Z]}]$.

\begin{figure}[t!]
  \declrel{Instantiation (outer)}
  {$\baseInstOut[\Delta_{\text{in}}]{\type_{\text{in}}}{\overline\EArgs_{\text{in}}}{\typec{\overline U}_{\text{out}}}{\type_{\text{out}}}$}
  \begin{gather*}
    \infrule{\ruleIOut}
    {\baseInstIn {\typeI 1} {\overline\EArgs}\Theta{\typec{\overline U}}{\typeI 2}}
            {\baseInstOut {\typeI 1}{\overline\EArgs}{\typec{\overline U}}{\typeI 2}}
  \end{gather*}
  \declrel{Instantiation (inner)}
  {$\baseInstIn[\Delta_{\text{in}}]{\type_{\text{in}}}{\overline\EArgs_{\text{in}}}{\Theta_{\text{out}}}{\typec{\overline U}_{\text{out}}}{\type_{\text{out}}}$}
  \begin{mathpar}
    \axiom{\ruleIRes}{\baseInstIn \type\epsilon\epsilon\epsilon\type}

    \infrule{\ruleIAll}
       {
         \IVar\fresh\
         \\
         \baseInstIn {\subs \IVar\TVar{\typeI 1}}{\expr,\overline\EArgs}\Theta{\typec{\overline U}}{\typeI 2}
       }
       {\baseInstIn {\TAll\TVar\kind{\typeI 1}}{\expr,\overline\EArgs}\Theta{\typec{\overline U}}{\typeI 2}}

       \infrule{\ruleITArg}
       { \baseInstIn {\subs{\typeI 2}\TVar{\typeI 1}}{\overline\EArgs}\Theta{\typec{\overline U}}{\typeI 3}
       }
       {\baseInstIn {\TAll\TVar\kind{\typeI 1}}{\typeI 2,\overline\EArgs}\Theta{\typec{\overline U}}{\typeI 3}}   

       \infrule{\ruleIArg}
       { \quicklook\expr{\typeI 1}{\Theta_1}
         \\
         \baseInstIn{\Theta_1\typeI 2}{\overline\EArgs}{\Theta_2}{\typec{\overline U}}{\typeI 3} \quad \Theta = \Theta_2 \circ \Theta_1
       }
       {\baseInstIn {\TFun{\typeI 1}{\typeI 2}}{\expr, \overline\EArgs}\Theta{\Theta{\typeI 1}, \typec{\overline U}}{\typeI 3}}   

       \infrule{\ruleIVar}
       { \IVar[Y],\IVar[Z]\fresh
         \\
         \baseInstIn {\TFun[\un]{\IVar[Y]}{\IVar[Z]}}{\expr, \overline\EArgs}{\Theta}{\typec{\overline U}}\type
       }
       {\baseInstIn {\IVar}{\expr, \overline\EArgs}{\{\IVar := \TFun[\un]{\IVar[Y]}{\IVar[Z]}\} \circ \Theta}{\typec{\overline U}}\type}   
     \end{mathpar}
     \declrel{Quick Look}{$\quicklook[\Gamma_{\text{in}}]{\expr_{\text{in}}}{\type_{\text{in}}}{\Theta_{\text{out}}}$}
  \begin{mathpar}
    \infrule{\ruleQLApp}
    {\isExprAlg[\Empty]\Gamma\EAppHead{\typeI 2}\Gamma
      \\
        \baseInstOut {\typeI 2}{\overline\EArgs}{\typec{\overline U}}{\typeI 3} }
      {\quicklook{\EApp\EAppHead{\overline\EArgs}}{\typeI 1}{
          \unify[\Empty]{\typeI 1}{\typeI 3}{\Theta}}}

      \infrule{\ruleQLOther}
       {\isExprAlg[\Empty]\Gamma\expr{\typeI 2}\Gamma}
       {\quicklook{\expr}{\IVar}{[\IVar := \typeI 2]}}
  \end{mathpar}
  \caption{Instantiation}
  \label{fig:base-inst}
\end{figure}


The final judgment, in \cref{fig:base-inst}, is used to investigate nested
applications and perform a possibly impredicative instantiation. Judgements of
the form
$\quicklook[\Gamma_{\text{in}}]{\expr_{\text{in}}}{\type_{\text{in}}}{\Theta_{\text{out}}}$
produce a substitution $\Theta$ from analyzing the expression $\expr$
with the expected type $\type$. The first rule, \ruleQLApp, handles nested
applications and type variables, synthesizing the application head $\EAppHead$
to obtain type $\typeI 2$, then matching $\typeI 1$ with the instantiation of
$\typeI 2$. In the remaining cases, \ruleQLOther produces a substitution with
the synthesized type.


All notions introduced so far are used in our \emph{bidirectional typing}
system. The bidirectional approach was introduced by Pierce and
Turner~\cite{PierceT98} as a two-way mechanism to propagate information: either
by ``pushing'' information down a type or synthetise it as usual. Our previous
work \cite{AlmeidaMTV22} uses the bidirectional approach in the algorithmic
typing, thus facilitating the implementation of local type inference. We
present alternative rules for the both directions of the application rule.

The judgements are now of form
$\isExprAlg[\Delta_{\text{in}}] {\Gamma_{\text{in}}} {\expr_{\text{in}}}
{\type_{\text{out}}} {\Gamma_{\text{out}}}$ for synthetizing a type and of form
$\isExprAlgAgainst [\Delta_{\text{in}}] {\Gamma_{\text{in}}} {\expr_{\text{in}}}
{\type_{\text{in}}} {\Gamma_{\text{out}}}$ for checking against a given type. A
difference in our rules regarding those of Serrano \etal\cite{SerranoHJV20} is
that we have to account for linearity therefore we use the entire context
$\Gamma$ when checking the first subexpression and pass the unused part to the
next subexpression \cite{walker:substructural-type-systems}. The rules are in \cref{fig:typing}.

We start with the synthesis rule \ruleSApp, which is slightly simpler since no
opportunity for matching is presented.
The rule first synthesises type $\typec{T_1}$ from head $\EAppHead$ of
type application. Then, parameters $\termc{\overline\EArgs}$ (types and
expressions) are instantiated to obtain the types $\typec{\overline U}$ of the
expressions in $\termc{\overline\EArgs}$ and the inferred type $\typec{T_2}$ of
the goal expression $\EApp\EAppHead{\overline\EArgs}$.
Function valargs extracts the expressions $\expr[e_1,\dots,e_n]$ in
$\termc{\overline\EArgs}$, discarding types. Each expression is then checked
against its expected type $\typec{U_i}$.
The initial typing context $\typec{\Gamma_1}$ is used check expression
$\termc{e_1}$, producing context $\typec{\Gamma_2}$, which is then passed to
check expression $\termc{e_2}$. The final context, $\typec{\Gamma_{n+1}}$ is
then the resulting context of rule \ruleCApp, together with type $\typec{T_2}$
produced by instantiation.

Rule \ruleCApp follows \ruleSApp until instantiation. The difference is that
here we have a type ($\typec{T_1}$) to match against.
The type of the expected type $\typec{T_1}$ is then matched against the
instantiated type $\typec{T_3}$ to obtain a substitution $\Theta$. At this point
we call type equivalence ($\TypeEquiv$) on types $\Theta{\typeI 1}$ and
$\Theta{\typeI 3}$.
To type the expressions in $\termc{\overline\EArgs}$ we proceed as in \ruleSApp,
only that we check each expression against $\Theta\typec{U_i}$.


\begin{figure}[t!]
  \declrel{Bidirectional typing}{
   $\isExprAlgAgainst[\Delta_{\text{in}}]{\Gamma_{\text{in}}}{\expr_{\text{in}}}{\type_{\text{in}}}{\Gamma_{\text{out}}}$
    and
    $\isExprAlg[\Delta_{\text{in}}]{\Gamma_{\text{in}}}{\expr_{\text{in}}}{\type_{\text{out}}}{\Gamma_{\text{out}}}$
  }
  \begin{mathpar}
    \infrule{\ruleSApp}
    {\isExprAlg{\Gamma_1}\EAppHead{\typeI 1}\Gamma_1
      \\
      \baseInstOut{\typeI 1} {\expr[\overline\EArgs]} {\typec{\overline U}}{\typeI 2}
      \\
      \expr[e_1,\dots,e_n] = \blk{\text{valargs}(}\expr[\overline\EArgs]\blk)
      \\
      \isExprAlgAgainst {\Gamma_i} {\expr[e_i]} {\typec{U_i}} {\Gamma_{i+1}}
      \\
      \forall i \in \{1,\dots,n\}
    }
    {\isExprAlg {\Gamma_1} {\EApp\EAppHead{\overline\EArgs}}
      {\typeI 2} {\Gamma_{n+1}}}

    \infrule{\ruleCApp}
    {\isExprAlg{\Gamma_1}\EAppHead{\typeI 2}\Gamma_1
      \\
      \baseInstOut[\Gamma_1]{\typeI 2} {\expr[\overline\EArgs]} {\typec{\overline U}}{\typeI 3}
      \\
      \unify[\Empty]{\typeI 1}{\typeI 3}{\Theta}
      \\
      \Theta{\typeI 1} \TypeEquiv \Theta{\typeI 3}
      \\
      \expr[e_1,\dots,e_n] = \blk{\text{valargs}(}\expr[\overline\EArgs]\blk)
      \\
      \isExprAlgAgainst {\Gamma_i} {\expr[e_i]} {\Theta\typec{U_i}} {\Gamma_{i+1}}
      \\
      \forall i \in \{1,\dots,n\}
    }
    {\isExprAlgAgainst{\Gamma_1}{\EApp\EAppHead{\overline\EArgs}}{\typeI 1}{\Gamma_{n+1}}}
  \end{mathpar}
  \caption{Bidirectional typing}
  \label{fig:typing}
\end{figure}



\section{Evaluation}
\label{sec:results}

We integrated our approach into the \freest interpreter and conducted an
experiment to assess the efficacy of the type inference algorithm.

The experiment involved eliminating all explicit type annotations from the
\freest source code (more than 10,000 lines of code), and evaluating whether the
algorithm could accurately reconstruct the omitted type information and verify
the program's correctness through type checking.

Before executing the experiment, we first tested the algorithm with all
annotations present to ensure its functionality in cases where type information
was explicitly provided (in function application arguments). In previous
iterations of FreeST, function type signatures were mandatory, meaning the
algorithm was only responsible for inferring types at application sites, a task
it successfully accomplished.

These results demonstrate the feasibility of type inference within FreeST.
Future enhancements could focus on optimizing the inference mechanism to further
minimize the need for annotations in lambda-bound variables while maintaining
the observed accuracy and performance.


\section{Related work}
\label{sec:related-work}


Sessions types were formerly proposed by Honda et.
al~\cite{DBLP:conf/concur/Honda93,DBLP:conf/esop/HondaVK98,DBLP:conf/parle/TakeuchiHK94} and
their theory is mature enough to see its core principles and ideas 
embodied in a book recently published~\cite{GayVasconcelos25}.

In the last few decades, there have been considerable effort on enhancing the
expressivity of session types in several dimensions, including object-oriented
programming \cite{GayGRV15} to web programming \cite{LindleyM17}, functional
programming \cite{GayV10,Vasconcelos11,VasconcelosRG04}, or programming with
exceptions \cite{FowlerLMD19}. Bono et al. \cite{BonoPT13} consider predicative
polymorphism for a session-oriented language very close the proposal of Thiemann
and Vasconcelos \cite{ThiemannV16}. Gay \cite{Gay08} introduced the concept of
bounded polymorphism for values transmitted over communication channels.

Thiemann and Vasconcelos \cite{ThiemannV16} proposed context-free session types,
taking advantage of a sequential composition operator and predicative
polymorphism, extended later to impredicative polymorphism~\cite{AlmeidaMTV22}.
This gain in expressivity came at the price of requiring type annotations in
every type application. The burden of type annotations was transposed to
\freest~\cite{freest}, the programming language with context-free session types
that constitutes the focus of this work. Other than kind and type annotations
for polymorphism, we require no more annotations for deciding type equivalence
and rely on the algorithm developed by Almeida \etal \cite{AlmeidaMV19}.
Padovani \cite{Padovani17,Padovani19} proposes a language that relies on
explicit annotations in the source code to split protocols thus ensuring a
structural alignment between programs and their types. His approach simplifies
type equivalence, but requires annotations.
Aagaard et al. \cite{AagaardHJK18} adapt the notion of context-free session
types from Thiemann and Vasconcelos \cite{ThiemannV16} to the applied
$\pi$-calculus. They established session fidelity by translating their calculus
into the psi-calculus and define type equivalence via session type bisimulation.


The related work on local type inference is vast. The most influential work in
this design space is the seminal paper of Pierce and Turner~\cite{PierceT98}
that defined the approach for local type inference by locally synthesising type
arguments and by bidirectionally propagating types. Their work is in the context
of $\text{F}_{<:}$ which differs from our setting, which is System F. Their
local constraint solver deals with subtyping constraints whereas we deal with a
different set of problems (brought by context-free session types).
Similarly to Pierce and Turner, Zhou and Oliveira~\cite{ZhaoO22} present a
variant of System F with top and bottom types and a restricted form of
subtyping. They propose a local type inference mechanism that infers predicative
instantiations, but requires the impredicative ones to be annotated.
HMF~\cite{Leijen08} extends the Hindley-Milner type inference system to support
first-class polymorphism. They only require annotations in polymorphic
parameters and ambiguous impredicative instantiations, which may not be
predicable for programmers. 
Boxy type inference, as introduced by Vytiniotis \etal~\cite{VytiniotisWJ06},
allows bidirectional propagation of type annotations whose propagation direction
is controlled by ``boxy types''. However, boxy type inference only guesses
monotypes. FreezeML~\cite{EmrichLSCC20} is an extension of ML where the
programmer may mark the locations where not to instantiate polymorphic types.
Similarly to HMF, type annotations are only required on lambda abstractions used
in a polymorphic fashion. Both boxy types and FreezeML introduce additional
constructs to System F types, something we tried to avoid.

The works closest to ours are Quick Look~\cite{SerranoHJV20} and spine-local
type inference~\cite{JenkinsS18}. We follow the first more closely. Quick Look is
a highly localised inference algorithm for impredicativity that is expressive
enough to handle System F and requires no extension to types. We deviate from
Quick Look on unification since we need to handle session types and in the
bidirectional typing rules because we need to handle resources linearly. Another
distinctive point is that we do not rely on standard contraint-based techniques
to infer non-impredicative instantiations. We try to infer everything from the
local assumptions at application sites; in this point we are closer to the
proposal of Jenkins and Stump as they infer everything from the application
spine~\cite{JenkinsS18}.

\section{Conclusion and future work}
\label{sec:future}
In this paper we propose a local type inference algorithm for \freest,
a programming language based on System F where processes communicate 
by message passing on channels governed by context-free session types.
We propose a bidirectional typing algorithm that takes advantage of 
a novel type matching algorithm. To properly infer type matching, we expicitly
handle the non-regular nature of recursion on context-free
session types and the monoidal laws introduced by the sequential 
composition operator.

The findings of this study confirm that type inference in \freest can
successfully reconstruct type information across our test suite that includes
tests that involve context-free session types, polymorphism-heavy tests
such as self-applications, polymorphic list encodings, and Church encodings.
These results underscore the effectiveness of our approach while also
highlighting areas for theoretical refinement. 

A key direction for future work is to establish formal guarantees for the
inference algorithm. 
The most immediate step is to prove the termination of type
matching and the correctness of the algorithm.
Another avenue for future work is to study an extension of the
proposed algorithm to infer annotations on priorities, as a way to ensure 
deadlock-freedom.
\nocite{*}

\paragraph{Acknowledgements}
\sloppy
Support for this research was provided by the Fundação para a Ciência e a
Tecnologia through project SafeSessions, ref. PTDC/CCI-COM/6453/2020, DOI
https://doi.org/10.54499/PTDC/CCI-COM/6453/2020, project REACT, ref.
2023.13752.PEX, DOI https://doi.org/10.54499/2023.13752.PEX, by the LASIGE
Research Unit, ref. UID/00408/2025, and by the COST Action CA20111.

\bibliographystyle{eptcs}
\bibliography{biblio}
\end{document}